\documentclass{article}
\usepackage{amsthm}
\usepackage{amsmath, amssymb, amscd}

\theoremstyle{plain}

\newtheorem{thm}{Theorem}[section]

\newtheorem{lem}{Lemma}[section]

\theoremstyle{definition}

\date{}
\numberwithin{equation}{section}
\begin{document}
\title{Nonspontaneous Supersymmetry Breaking}
\author{Alexander 
Golubev\footnote{e-mail address: \texttt{agolubev@nyit.edu}}\\
New York Institute of Technology, New York, NY 10023}
\maketitle
\begin{abstract}
A new way of supersymmetry breaking involving a dynamical
parameter is introduced. It is independent of particle 
phenomenology and gauge groups. The only requirement is that
Lorentz invariance be valid strictly infinitesimally (i. e.
Spin(1, 3) be for some values of the parameter replaced by
a compact group $G$ with its Lie algebra $ \mathfrak{g}
\cong \mathfrak{so}(1,3)$). 
\end{abstract}

\section{Introduction}
The particles predicted by supersymmetric field theories failed to 
appear in experiments, so that within the accessible range of energies
supersymmetry must be broken. However, if the Standard Model is to 
serve as a ($\thickapprox 300$ GeV) low-energy approximation to some
as-yet-unknown unified theory, supersymmetry would have to manifest at
a higher energy level to allow supermultiplets involving scalar
particles to be formed, thus preventing  those scalar particles
from acquiring large ($\thickapprox 10^{16} -10^{18}$ GeV) bare masses 
and bringing corpulence to the presently observable particles~\cite{W}. 
This disparity of characteristic mass scales, known as the hierarchy 
problem, and the above-outlined way around it
provide the strongest theoretical motivation to keep 
supersymmetry alive. Another feature of supersymmetric theories that
is considered desirable is the presence of sparticles - superpartners
of the observable ones. They are natural candidates for exotica such
as the missing `dark matter' of the Universe.\\
\indent
Without exception, all mechanisms of supersymmetry breaking hitherto
proposed are spontaneous. Generally speaking, spontaneous supersymmetry
breaking occurs when the variation of some field under supersymmetry
transformations yields nonzero vacuum expectation values:
\begin{equation*}
\langle 0|\delta (\text{field})|0 \rangle \ne 0.
\end{equation*}
\noindent
As a result, the vacuum state gains energy, and enters supermultiplets
opposite a massless fermion - the goldstino. If gravitation is present,
supersymmetry localizes and instead of the goldstino, the gravitino
becomes the vacuum's superpartner.\\
\indent
Unlike the breaking of electroweak symmetry, a direct coupling
of the electroweak force to the resulting Higgs particles is not
possible because such a coupling leads to sum rules for the masses
of the unobserved superpartners that are excluded. An indirect
transmission of supersymmtry breaking to the observable sector is
needed. Based on the particulars of coupling, the spontaneous 
supersymmetry breaking mechanisms are: gauge-mediated (GMSB)
~\cite{G-R}, anomaly-mediated (AMSB)~\cite{L-S}, supergravity
(SUGRA)~\cite{B-W}, and extra-dimensional~\cite{R-S}. These 
different mechanisms have characteristic mass spectra and 
experimental signatures.\\
\indent
The search for sparticles goes on, but the experimental data
recently obtained at LEP and Tevatron~\cite{Nnn} does not
encourage optimism on the subject of plausibility of GMSB and
AMSB. SUGRA is unassailable, although its superpotential contains a 
soft parameter chosen to fit the experimentally confirmed phenomena.
That raises the question whether there is a viable alternative to the
brane theory treatment of the hierarchy problem in particular and to
spontaneous supersymmetry breaking in general.\\ 
\indent
Our paper is an attempt to explain the why of the hierarchy problem
in terms of fundamental space-time symmetries. As far as we know,
the first results linking supersymmetry algebras to space-time
symmetries were published by Nahm~\cite{N}. The anti-de-Sitter
space with its $O(3,2)$ symmetries supports all conceivable
supersymmetry algebras, whereas the de-Sitter space having
$O(4,1)$ as the symmetry group has only $N=2$ supersymmetry.
Thus this Universe evolving from the anti-de-Sitter to the 
de-Sitter regime may provide a toy model of nonspontaneous
$N \ne 2$ supersymmetry breaking. We use the word `nonspontaneous'
advisedly, for no goldstino (or gravitino) is created. It is
very instructive to expose the fatal flaw of this model. There is
no smooth direct parametric transition from $O(3,2)$ to $O(4,1)$
because $\mathfrak{o}(3,2) \ncong \mathfrak{o}(4,1)$, and for some
value of the parameter space-time symmetries collapse even
infinitesimally.\\
\indent
Therefore to make such a theory work one needs a family of locally
isomorphic Lie groups, smoothly depending on a parameter, and
differing in their facility to support supersymmetry algebras. Then
the parameter may be interpreted as the energy scale, pre- and
post-unification values separated by an interval. In the Minkowski
${\mathbb{R}}^4$ one also requires Lorentz invariance. That could
only be satisfied for families of Lie groups locally isomorphic to
the Lorentz group, and containing that group as a member. In what
follows we find one such family containing, at one extreme Spin(1, 3),
and at the other a compact Lie group $G$ which, while allowing
to maintain Lorentz invariance, does not support any supersymmetry
algebras. Then the parametric evolution from the former to the 
latter constitutes a mechanism of supersymmetry breaking. The
rationale behind our construction is deceptively simple: 
supersymmetry happens to be broken (or, rather, nonexistent) below
the unification mark because the respective $S$-matrix has 
finite-dimensional blocks - exacting finite-dimensional unitary 
representations (i. e. compactness) 
of the (respective) symmetry group. By contrast,
above the unification mark more off-diagonal elements of the 
$S$-matrix become non-zero; consequently, the erstwhile 
finite-dimensional blocks coalesce, the representation spaces 
become infinite-dimensional, and Spin(1,3) takes over.\\
\indent
There are experimentally verifiable effects associated with
the symmetry group $G$. The electromagnetic vector potentials
transform differently under Spin(1,3), and that difference 
can be detected.\\
\indent
Lastly, we dispense with the physical constants by setting
$ \hbar \;=\;c\;=\;1.$

\section{Mathematical Preliminaries}

The Pauli matrices are
\begin{equation}
{\sigma}_1=
\begin{bmatrix}
0&1\\
1&0
\end{bmatrix},\; {\sigma}_2=
\begin{bmatrix}
0&-i\\
i&\phantom{-}0
\end{bmatrix},\; {\sigma}_3=
\begin{bmatrix}
1&\phantom{-}0 \\
0&-1
\end{bmatrix}.
\end{equation}
\indent
The Dirac representation of $SU(2)$, denoted
$SU_{\mathcal{D}}(2)$ is generated by
\begin{equation}
J_1=\frac{1}{2}
\begin{bmatrix}
{\sigma}_1&0\\
0&{\sigma}_1
\end{bmatrix},\; J_2=\frac{1}{2}
\begin{bmatrix}
{\sigma}_2&0\\
0&{\sigma}_2
\end{bmatrix},\; J_3=\frac{1}{2}
\begin{bmatrix}
{\sigma}_3&0\\
0&{\sigma}_3
\end{bmatrix}.
\end{equation}
\noindent
There still exists the twofold covering epimorphism of Lie groups:
\begin{equation}
\mathcal{A}: \;\;SU_{\mathcal{D}}(2) \longrightarrow 
\begin{bmatrix}
[SO(3)]&0\\
0&1
\end{bmatrix}.
\end{equation}
\noindent
Spin(1,3) may be viewed as a complex extension of $SU_{\mathcal{D}}(2)$:
\begin{equation}\label{Lorentzext}
\left\{ J_i = \frac{1}{2}
\begin{bmatrix}
{\sigma}_i&0\\
0&{\sigma}_i
\end{bmatrix} \right\} \mapsto
\left\{ J_i =\frac{1}{2}
\begin{bmatrix}
{\sigma}_i&0\\
0&{\sigma}_i
\end{bmatrix},\; K^{\mathbb{C}}_i=\frac{1}{2}
\begin{bmatrix}
i&\phantom{-}0 \\ 
0&-i
\end{bmatrix}
\begin{bmatrix}
{\sigma}_i&0\\
0&{\sigma}_i
\end{bmatrix}\right\}.
\end{equation}
\noindent
Fortuitously, there is a  class of mutually isomorphic almost complex 
Lie algebra extensions,
of which $\mathfrak{so}(1,3)$, generated by $\{J_i,K^{\mathbb{C}}_i\}$ 
of \eqref{Lorentzext} is a member. We are interested mainly in the
following almost complex extension:
\begin{equation}\label{Gext}
\left\{ J_i = \frac{1}{2}
\begin{bmatrix}
{\sigma}_i&0\\
0&{\sigma}_i
\end{bmatrix} \right\} \mapsto
\left\{ J_i =\frac{1}{2}
\begin{bmatrix}
{\sigma}_i&0\\
0&{\sigma}_i
\end{bmatrix},\; K_i=\frac{1}{2}
\begin{bmatrix}
\phantom{-}0&1 \\ 
-1&0
\end{bmatrix}
\begin{bmatrix}
{\sigma}_i&0\\
0&{\sigma}_i
\end{bmatrix}\right\}.
\end{equation}
\noindent
Its relevant properties are summarized in
\begin{thm}\label{Gcompactness}
There exists a unique compact semisimple Lie group $G \subset SU(4)$,
whose Lie algebra $ \mathfrak{g}\cong \mathfrak{so}(1,3)$ is generated
by \eqref{Gext}.
\end{thm}
\begin{proof}
Every almost complex extension corresponds (up to a nonzero factor)
to a matrix
\begin{equation*}
\begin{bmatrix}
a&b\\
c&d
\end{bmatrix}\in U(2),\;\;
\begin{bmatrix}
a&b\\
c&d
\end{bmatrix}
\begin{bmatrix}
a&b\\
c&d
\end{bmatrix}=
\begin{bmatrix}
-1&\phantom{-}0 \\ 
\phantom{-}0&-1
\end{bmatrix}.
\end{equation*}
\noindent
$ \mathfrak{g}\cong \mathfrak{so}(1,3)$ implies $ad -bc=1$.
Therefore
\begin{equation*}
\Re a = \Re d=0,\; \Im c = \Im b, \; \Re c =- \Re b.
\end{equation*}
\noindent
This allows us to write the most general almost complex
extension as
\begin{equation*}
J_i \mapsto \left(w
\begin{bmatrix}
i&\phantom{-}0 \\ 
0&-i
\end{bmatrix} + u
\begin{bmatrix}
0&i \\ 
i&0
\end{bmatrix} + v
\begin{bmatrix}
\phantom{-}0&1\\
-1&0
\end{bmatrix}\right)J_i,\; w^2 + u^2 + v^2 =1.
\end{equation*}
\indent
To ensure compactness, we must have $\exp i{\kappa}^aK_a$ bounded. 
Whence $w=0$, $u=0$ is the only choice. And this is \eqref{Gext}.\\
\indent
According to Helgason (\cite{He}, Chapter~II, \S 2, Theorem~2.1), 
there exists a Lie group $G$, whose Lie algebra is generated by 
$\{J_i, K_i\}$ of \eqref{Gext}. Its elements are all of the form
$\exp i({\theta}^bJ_b + {\kappa}^a K_a)$, which means $G$ is a 
Lie subgroup of $SU(4)$.
Now $G$ has to be closed in the standard matrix topology of $SU(4)$.
That is based on a fundamental
result of Mostow~\cite{M}: any semisimple Lie subgroup $H$ of a compact 
Lie group $C$ is closed in the relative topology of $C$. In our case,
$SU(4)$ is compact, $\mathfrak{g}$ is semisimple.
\end{proof}
\indent
In the sequel we will work with the homogeneous space 
$G/SU_{\mathcal{D}}(2)$.
\begin{lem}\label{trivialhomotopy}
\begin{equation*}
{\pi}_1(G/SU_{\mathcal{D}}(2))= 0.
\end{equation*}
\end{lem}
\begin{proof}
For all Lie groups ${\pi}_2 (.) = 0$~\cite{B}; for 
$SU_{\mathcal{D}}(2)$, 
${\pi}_0(SU_{\mathcal{D}}(2))= 0$ by connectedness. Also, 
$SU_{\mathcal{D}}(2)$ is a
closed subgroup of $SU(4)$ in the ordinary matrix topology.
We therefore have the following exact homotopy sequence~\cite{B}:
\begin{equation*}
\begin{split}
0 \rightarrow {\pi}_2 (SU(4)/SU_{\mathcal{D}}(2))& 
\rightarrow {\pi}_1 (SU_{\mathcal{D}}(2))\\
&\rightarrow {\pi}_1 (SU(4)) \rightarrow {\pi}_1 
(SU(4)/SU_{\mathcal{D}}(2))\rightarrow 0.
\end{split}
\end{equation*}
\noindent
${\pi}_1 (SU(4))= 0$~\cite{B} whence 
\begin{equation*}
{\pi}_1 (SU(4)/SU_{\mathcal{D}}(2)) \cong {\pi}_1 
(SU_{\mathcal{D}}(2)) = {\pi}_1({\mathbb{S}}^3) =0.
\end{equation*}
\noindent
Now homotopy is functorial. The embedding
$\xi : G/SU_{\mathcal{D}}(2) \hookrightarrow 
SU(4)/SU_{\mathcal{D}}(2)$ 
induces the monomorphism of fundamental groups
\begin{equation*}
{\xi}_{\pi *} : {\pi}_1 (G/SU_{\mathcal{D}}(2)) \rightarrow
{\pi}_1 (SU(4)/SU_{\mathcal{D}}(2)). \qedhere
\end{equation*}
\end{proof}
\begin{thm}
\begin{equation*}\label{s3theorem}
 G/SU_{\mathcal{D}}(2) \cong {\mathbb{S}}^3.
\end{equation*}
\end{thm}
\begin{proof}
\noindent
$\mathfrak{g}$ 
decomposes as a vector space into two three-dimensional
subspaces,
\begin{equation*}
\mathfrak{g} = \mathfrak{j} \oplus \mathfrak{k},
\end{equation*}
\noindent
Based on this decomposition, there is an involutive automorphism 
\begin{equation*}
\vartheta : \mathfrak{g} \;
\longrightarrow \;
\mathfrak{g} 
\end{equation*}
\noindent
defined by
\begin{equation*}
\vartheta (J + K) = J - K,\quad \forall J \in \mathfrak{j},
\quad \forall K \in \mathfrak{k}.
\end{equation*}
\noindent
$\mathfrak{j}$ is the set of fixed points of $\vartheta$. It is 
unique (\cite{He}, Chapter IV, \S 3, Proposition 3.5). The pair 
$(\mathfrak{g} \;,\; \vartheta)$ is an orthogonal symmetric Lie algebra
(\cite{He}, Chapter~IV, \S 3). There is a Riemannian symmetric
pair $(G,\; SU_{\mathcal{D}}(2))$ associated with $(\mathfrak{g}, \; 
 \vartheta)$ so that the quotient
$G/SU_{\mathcal{D}} (2)$ is a complete locally symmetric Riemannian space.
Furthermore, its curvature corresponding to any $ G$-invariant
Riemannian structure is given by (\cite{He}, Chapter IV, \S 4, 
Theorem 4.2):
\begin{equation*}
R(K_{i_1}, K_{i_2})K_{i_3}= -[[K_{i_1}, K_{i_2}], K_{i_3}]
\quad \forall K_{i_1}, K_{i_2}, K_{i_3} \in \mathfrak{k}.
\end{equation*}
\noindent
Computing the sectional curvature we see that $R^{\text{sect}} 
\equiv 1$. Now a pedestrian version of the Sphere 
theorem~\cite{C-G} asseverates that a complete simply connected 
Riemannian manifold with $R^{\text{sect}} \equiv 1$ is isometric 
to a sphere of appropriate dimension. In our case the topological
condition is satisfied in view of Lemma~\ref{trivialhomotopy}.
\end{proof}
\indent
Consider the natural inclusions of Lie groups
\begin{equation}
\iota : G \hookrightarrow GL(4, \mathbb{C}), \quad
\iota : \text{Spin}(1, 3) \hookrightarrow GL(4, \mathbb{C}).
\end{equation}
\noindent
Their images inside $GL(4, \mathbb{C})$ intersect:
\begin{equation}\label{properspin}
\iota (G) \cap \iota (\text{Spin}(1, 3))=SU_{\mathcal{D}}(2).
\end{equation} 
\noindent
Because of~\eqref{properspin}, the set 
\begin{equation}
{\text{Ad}}_{\iota (G)}(\iota (\text{Spin}(1, 3)))=
\coprod_{U \in G} U\text{Spin}(1, 3)U^H, 
\end{equation}
the disjoint union of conjugates of $\text{Spin}(1, 3))$,
has the same cardinality as the set of 
all boosts in $G$. Similarly, there is the natural inclusion
\begin{equation}
\iota :\quad SO(4) \hookrightarrow GL(4, \mathbb{R}).
\end{equation}
\noindent
The set ${\text{Ad}}_{\iota (SO(4))}(\iota (\text{SO}(1, 3)^e))$ is
homeomorphic to $SO(4)/SO(3) \cong {\mathbb{S}}^3$. Combining this with
Theorem~\ref{s3theorem} we arrive at:
\begin{equation}\label{wpdiff}
\begin{CD}
\text{Ad}_{\iota (G)}(\iota (\text{Spin}(1,3))) @=
G/SU_{\mathcal{D}} (2) @>{\cong}>> {\mathbb{S}}^3\\ 
@.             @V{\wp}VV                 @|\\
\text{Ad}_{\iota (SO(4))}(\iota (\text{SO}(1,3)^e))
@= SO(4)/SO(3) @>{\cong}>>  {\mathbb{S}}^3
\end{CD}
\end{equation}
\noindent
The double horizontal lines indicate set-theoretic bijective 
correspondences, the upper $\cong$ is an isometry, the lower one 
is a diffeomorphism. Furthermore, the diagram~\eqref{wpdiff} 
commutes and \textit{de facto} defines the diffeomorphism $\wp$.
This diffeomorpism is utilized in the sequel to effect an
action of $G$.

\section{Equivariant Impulse Operators}

\indent
$G$ does not act on the Minkowski ${\mathbb{R}}^4$ by isometries.
We have
\begin{equation}
\begin{cases}
G \times {\mathbb{R}}^4 \longrightarrow {\mathbb{R}}^4;\\
(\exp i({\theta}^b J_b +{\kappa}^a K_a), x^{\mu}) \mapsto
x'^{\mu}= {\mathcal{A}(\exp i{\theta}^b J_b)}^{\mu}_{\lambda}
{\wp(\exp i{\kappa}^a K_a)}^{\lambda}_{\eta} x^{\eta}.
\end{cases}
\end{equation}
\noindent
In fact, the metric becomes frame-dependent:
\begin{equation*}
\wp (\exp i{\kappa}^a (\alpha) K_a) g \wp (\exp(- i{\kappa}^a (\alpha) K_a))
\ne g, \; \alpha \in [0, 2\pi],\; \alpha \ne \{0, 2\pi \}; 
\end{equation*}
$\alpha$ being the group parameter here. Yet physical
quantities must remain frame-independent. Therefore,
instead of the standard quantum field theory substitution
\begin{equation}\label{standardQFT}
P_{\mu} \longrightarrow i{\partial}_{\mu},
\end{equation}
\noindent
we employ the rule
\begin{equation}\label{nabla}
P_{\mu} \longrightarrow i{\nabla}_{\mu}({\alpha}) 
\overset{\text{def}}{=} i({\varepsilon}^{\nu}_{\mu}(\alpha)
{\partial}_{\nu} + i{\kappa}_{\mu}^a (\alpha) K_a),
\end{equation}
\noindent
the exact form of ${\varepsilon}^{\nu}_{\mu}(\alpha)$ and
${\kappa}_{\mu}^a (\alpha)$ to be determined. $K_a$'s are
in keeping with the (1/2, 1/2) representation of $P_{\mu}$'s.
This construction is an equivariant incarnation of the free
spin structure due to Plymen and Westbury~\cite{P-W}. 
Briefly, let $M$ be a 4-dimensional 
smooth manifold with all the obstructions to the existence of 
a Lorentzian metric vanishing (for instance, a parallelilazable 
$M$ would do). Let 
\begin{equation*}
\Lambda : \;\; \textnormal{Spin} (1, 3) \rightarrow SO(1, 3)^e
\end{equation*}
be the twofold covering epimorphism of Lie groups.
A free spin structure on $M$ consists of a principal bundle 
$\zeta : \Sigma \rightarrow M$ with
structure group $\textnormal{Spin} (1, 3)$ and a bundle map
$\widetilde{\Lambda}: \Sigma \rightarrow \mathcal{F} M$ into the bundle
of linear frames for $TM$, such that
\begin{equation*}
\widetilde{\Lambda} \circ {\widetilde{R}}_S = 
{\widetilde{R}}'_{\iota \circ \Lambda (S)} \circ \widetilde{\Lambda}
\;\;\; \forall S \in \textnormal{Spin} (1, 3),
\end{equation*}
\begin{equation*}
{\zeta}' \circ \widetilde{\Lambda} = \zeta,
\end{equation*}
${\widetilde{R}}$ and ${\widetilde{R}}'$ being the canonical right actions
on $\Sigma$ and $\mathcal{F} M$ respectively, $\iota : SO(1,3)^e \rightarrow
GL(4, \mathbb{R})$ the natural inclusion of Lie groups, and ${\pi}' :
\mathcal{F} M \rightarrow M$ the canonical projection. The map 
$\widetilde{\Lambda}$ is called a spin-frame on $\textnormal{Spin}(1, 3)$.
This definition of a spin structure induces metrics on $\Sigma$. Indeed,
given a spin-frame $\widetilde{\Lambda}: \Sigma \rightarrow \mathcal{F} M$,
a dynamic metric $g_{\widetilde{\Lambda}}$ is defined to
be the metric that ensures orthonormality of all frames in 
$\widetilde{\Lambda}(\Sigma) \subset \mathcal{F} M$. It should be
emphasized that within the Plymen and Westbury's formalism the 
metrics are built \textit{a posteriori},
after a spin-frame has been set by the field equations. In our
formalism the metrics are obtained via the $G$-action, and the
set of all allowable metrics is $\text{Ad}_{\iota (SO(4))}
(\iota (\text{SO}(1,3)^e))$.\\
\indent
${\nabla}_{\mu}({\alpha})$ qualifies as a $G$-connection on the 
principal $G$-bundle over the physical space-time. Furthermore,
we impose an additional condition on~\eqref{nabla} to ensure
validity of the relativistic impulse-energy identity:
\begin{equation}\label{coinvariance}
{P}^{\mu}(\alpha){P}_{\mu}(\alpha)= g^{\nu \lambda}({\alpha})
{\nabla}_{\nu}({\alpha}){\nabla}_{\lambda}({\alpha})
\overset{\text{def}}{=} g^{\nu \lambda}(0)
{\partial}_{\nu}{\partial}_{\lambda}= P^{\mu}(0)P_{\mu}(0).
\end{equation}
\noindent
This translates to some algebraic relations among ${\kappa}^a_{\mu}$'s
and ${\varepsilon}^{\nu}_{\mu}$'s. However, we still need to make
the $G$-transformation law of~\eqref{nabla} more explicit. First,
these operators are natural spinors in the sense that $SU_{\mathcal{D}}(2)$
acts linearly:
\begin{alignat}{2}\label{su2action}
U{\gamma}^{\mu}{\nabla}_{\mu}U^H &=U{\gamma}^{\mu}U^H
{\varepsilon}^{\nu}_{\mu}{\partial}_{\nu} + i{\kappa}_{{\mu}}^a
U{\gamma}^{\mu}U^HUK_aU^H& &\quad\\
&=M^{\mu}_{\eta}{\gamma}^{\eta}{\varepsilon}^{\nu}_{\mu}{\partial}_{\nu}
+M^{\mu}_{\eta}{\gamma}^{\eta}i{\kappa}_{{\mu}}^a r^n_a K_n & &
\quad \text{by}\;[\mathfrak{j}, \mathfrak{k}] = \mathfrak{k}.\notag
\end{alignat}
\noindent
Here $M^{\mu}_{\eta}$'s realize an $SO(3)$ transformation 
$(U \in SU_{\mathcal{D}}(2))$, which is
at its most transparent if ${\gamma}^{0}$ is diagonal. 
As for $r_a^n$'s, they determine how the potentials behave:
\begin{equation}
{\Tilde{\kappa}}^a_{\mu}={\kappa}_{{\mu}}^1 r^a_1 +
{\kappa}_{{\mu}}^2 r^a_2 + {\kappa}_{{\mu}}^3 r^a_3, \quad {\text{and}}
\end{equation}
\begin{equation}
{|r^a_1|}^2 +{|r^a_2|}^2 +{|r^a_3|}^2=1,\quad a=\{1,2,3\}.
\end{equation} 
\noindent
To see how they are boosted, we treat a prototypical case - that of a
boost in the $x^{3}$-direction. Specifically,
\begin{align}
{\nabla}_{0}&={\varepsilon}^{0}_{0}(\alpha){\partial}_0 + 
{\varepsilon}^{3}_{0}(\alpha){\partial}_3 + i{\kappa}_0(\alpha)K_3,\\
{\nabla}_{3}&={\varepsilon}^{0}_{3}(\alpha){\partial}_0 + 
{\varepsilon}^{3}_{3}(\alpha){\partial}_3 + i{\kappa}_3(\alpha)K_3,\\
{\nabla}_{1} &= {\partial}_1,\\
{\nabla}_{2} &= {\partial}_2.
\end{align}
\noindent
We look for solutions of 
\begin{equation}\label{Dirac1}
(i {\gamma}^{\mu}{\nabla}_{\mu} - m)\Psi = 0,
\end{equation}
\noindent
modelled on the free spinors
\begin{equation}
\Psi(\alpha)= s(\alpha)e^{-i(p_0x^0 +p_3x^3)},
\end{equation}
subject to the relativistic impulse condition ${p_0}^2-{p_3}^2=m^2$.
In the standard representation
\begin{equation}
{\gamma}^0 =
\begin{bmatrix}
1&\phantom{-}0\\
0&-1
\end{bmatrix},\quad {\gamma}^{i} =
\begin{bmatrix}
0&-{\sigma}_i\\
{\sigma}_i&\phantom{-}0
\end{bmatrix},
\end{equation}
\noindent
the equation \eqref{Dirac1} yields the following matrix:
\begin{equation}
\begin{bmatrix}
{\varepsilon}_{0}(\alpha)-m(\alpha)
&0&-{\varepsilon}_{3}(\alpha)-{\kappa}_{0}(\alpha)&0\\
0&{\varepsilon}_{0}(\alpha)-m(\alpha)
&0&{\varepsilon}_{3}(\alpha)+{\kappa}_{0}(\alpha)\\
{\varepsilon}_{3}(\alpha)-{\kappa}_{0}(\alpha)
&0&-{\varepsilon}_{0}(\alpha)-m(\alpha)&0\\
0&-{\varepsilon}_{3}(\alpha)+{\kappa}_{0}(\alpha)
& 0&-{\varepsilon}_{0}(\alpha)-m (\alpha) 
\end{bmatrix}, 
\end{equation}
where the entries are
\begin{align}
{\varepsilon}_{0}(\alpha)&={\varepsilon}^{0}_{0}(\alpha)p_0 + 
{\varepsilon}^{3}_{0}(\alpha)p_3,\\
{\varepsilon}_{3}(\alpha)&={\varepsilon}^{0}_{3}(\alpha)p_0 + 
{\varepsilon}^{3}_{3}(\alpha)p_3,\\
m(\alpha)&= m +{\kappa}_{3} (\alpha).  
\end{align}
\noindent
Its rank has to be 2 for all values of $\alpha$, thus 
constraining ${\kappa}_0(\alpha)$ and ${\kappa}_3(\alpha)$:
\begin{equation}
{{\varepsilon}_{0}}^2(\alpha) -  {{\varepsilon}_{3}}^2(\alpha)=
(m +{\kappa}_{3}(\alpha))^2 -{{\kappa}_0}^2(\alpha).
\end{equation}
\noindent
Evidently ${\kappa}_{\mu}^a(\alpha)$'s are not identically zero. At the
same time, ${\kappa}_{\mu}^a(0) =0, \;\forall \mu = \{0,1,2,3\}$.
Hence, a boost entails a nonlinear change in the potentials.\\
\indent
Finally, we are in a position to deal with supersymmetry algebras. For
the reminder of this section, the impulse operators and all other
quantities expressly depend on the parameters introduced in the proof
of Theorem~\ref{Gcompactness}. For convenience, we bundle them into
one complex parameter $z$ via stereographic projection, so that
$K_a(0)= K_a^{\mathbb{C}}$, $K_a(1)=K_a$, ${\varepsilon}^{\nu}_{\mu}
(\alpha, 0)={\delta}^{\nu}_{\mu}$, ${\kappa}_{\mu}^a (\alpha, 0)=0$.  
Should there exist such algebras,
$ Q_m(z), {\bar{Q}}_m(z)$ would generate them. But they 
realize a linear representation of the (respective) symmetry group, and
we arrive at an equality impossible for some $z \in [0,1]$:
\begin{equation}
\{ Q_m(z), {\bar{Q}}_m(z) \} = 
-2i{\gamma}^{\mu}({\varepsilon}^{\nu}_{\mu}(\alpha, z)
{\partial}_{\nu} + i{\kappa}_{\mu}^a (\alpha, z) K_a(z)).
\end{equation}
\noindent
The right-hand side transforms nonlinearly because of 
${\kappa}_{\mu}^a (\alpha,1)$, whereby proving that there are
no $ Q_m(1), {\bar{Q}}_m(1)$. Adding central charges $Z_m$,
$Z^*_m$ on the right-hand side would not remedy the situation 
because these charges commute with the symmetry group generators.

\section{The Relativistic Aharonov-Bohm Effect}

\indent
The diffeomorphism between $G/SU_{\mathcal{D}}(2)$ and $SO(4)/SO(3)$
established in~\eqref{wpdiff} induces a vector space isomorphism 
${\wp}^*$, taking $K_i$'s into the matrices
\begin{equation*}
\begin{bmatrix}
\phantom{-}0&\phantom{-}0&\phantom{-}0&\phantom{-}i\\
\phantom{-}0&\phantom{-}0&\phantom{-}0&\phantom{-}0\\
\phantom{-}0&\phantom{-}0&\phantom{-}0&\phantom{-}0\\
-i         &\phantom{-}0&\phantom{-}0&\phantom{-}0
\end{bmatrix},\;
\begin{bmatrix}
\phantom{-}0&\phantom{-}0&\phantom{-}0&\phantom{-}0\\
\phantom{-}0&\phantom{-}0&\phantom{-}0&\phantom{-}i\\
\phantom{-}0&\phantom{-}0&\phantom{-}0&\phantom{-}0\\
\phantom{-}0&          -i&\phantom{-}0&\phantom{-}0
\end{bmatrix},\;
\begin{bmatrix}
\phantom{-}0&\phantom{-}0&\phantom{-}0&\phantom{-}0\\
\phantom{-}0&\phantom{-}0&\phantom{-}0&\phantom{-}0\\
\phantom{-}0&\phantom{-}0&\phantom{-}0&\phantom{-}i\\
\phantom{-}0&\phantom{-}0&        -i  &\phantom{-}0
\end{bmatrix},
\end{equation*}
\noindent
forming a subspace of $\mathfrak{so}(4)$. ${\wp}^*$ comes in handy
for differentiating vector potentials. Our operators ${\nabla}_{\mu}$
become
\begin{equation}
{\nabla}_{\mu}^{\wp} 
\overset{\text{def}}{=} {\varepsilon}^{\nu}_{\mu}
{\partial}_{\nu} + i{\kappa}_{\mu}^a {\wp}^*( K_a).
\end{equation}
\noindent
Consequently, ${\mathbb{A}}_{\mu}$'s transform via
${\mathbb{A}}_{\mu} \mapsto {\mathbb{A}}'_{\mu}$ such that
\begin{equation}\label{vectorpot}
{\nabla}_{\mu}^{\wp} {\mathbb{A}}'_{\nu}=
{\partial}_{\mu} {\Lambda}^{\eta}_{\nu}{\mathbb{A}}_{\eta}.
\end{equation}
\noindent
Here ${\Lambda}^{\eta}_{\nu}$ designates a pure Lorentz boost
in the direction determined by ${\kappa}_{\mu}^a K_a$.
Clearly, \eqref{vectorpot} is the only possible way to maintain
the Lorentz covariance of the electromagnetic field. But the
behavior of ${\mathbb{A}}_{\mu}$'s is not subject to any other
constraints. There are plenty of vector potentials that satisfy
\eqref{vectorpot}, yet enjoy some freedom: ${\mathbb{A}}'_{\mu}
\ne {\Lambda}^{\eta}_{\mu}{\mathbb{A}}_{\eta}$.\\
\indent
Now consider the setting of the Aharonov-Bohm experiment~\cite{A-B}.
If performed on the ground and on the aircraft moving fast enough
to make time slowing detectable, the difference in phase shifts
$(\Delta \varphi)_{v=0} -(\Delta \varphi)_{v \ne 0}$ compared 
with the theoretical values computed using the two transformation 
laws ($G$ \textit{versus} Spin(1,3)) would settle the question of which 
law better describes Nature within the given energy range.

\end{document}